\documentclass[a4paper,10pt,3p,onecolumn,preprint,fleqn,superscriptaddress,amsmath,amssymb,prl,floatfix]{article}

\usepackage{booktabs}
\usepackage[maxfloats=36]{morefloats}
\usepackage{array,multirow}
\usepackage{subfigure}
\usepackage{hyperref}
\usepackage{color}
\usepackage{pict2e}
\usepackage{tikz}
\usetikzlibrary{er}
\usepgflibrary{shapes}
\usepackage{pdfpages}
\usepackage{longtable}
\usepackage{float}
\restylefloat{table}
\usepackage{mathtools}
\usepackage{pst-node,pst-dbicons}
\usepackage{mdframed}
\usepackage{enumerate}
\usepackage[utf8]{inputenc}

\usetikzlibrary{shapes.multipart}

\usepackage{graphicx}
\usepackage{xcolor,colortbl}
\usepackage{graphics}
\usepackage{cite}
\usepackage{lastpage}
\usepackage{fancyhdr}
\usepackage{array}
\usepackage{lipsum}

\usepackage{tikz}
\usepackage{footmisc}
\usepackage{mathtools}
\usetikzlibrary{shapes.geometric, arrows}
\usepackage{blindtext}
\usepackage{shapepar}

\usepackage{hyperref}
\usepackage{authblk}

\usepackage{fancyhdr}

\usepackage{amsmath}
\newtheorem{lem}{\bf \textit{Lemma}}

\newtheorem{exa}{\bf \textit{Example}}

\newtheorem{theo}{\bf \textit{Theorem}}

\begin{document}
  
  
\title{Improved Upper Bound on Independent Domination Number for Hypercubes}

\author{Debabani Chowdhury \& Debesh K. Das}
\affil{Department of Computer Sc. \& Engg. \\Jadavpur University, Kolkata - 700 032, India\\\it{debabani.chowdhury@gmail.com}\\\it{debeshkdas@gmail.com}}
\author{Bhargab B. Bhattacharya}
\affil{Department of Computer Sc. \& Engg. \\Indian Institute of Technology Kharagpur\\ WB - 721 302, India\\\it{bbbiitkgp@gmail.com}}



\maketitle

\begin{abstract}
We revisit the problem of determining the independent domination number in hypercubes for which the known upper bound is still not tight for general dimensions. We present here a constructive method to build an independent dominating set $S_n$ for the $n$-dimensional hypercube $Q_n$, where $n=2p+1$, $p$ being a positive integer $\ge 1$, provided an independent dominating set $S_p$ for the $p$-dimensional hypercube $Q_p$, is known. The procedure also computes the minimum independent dominating set for all $n=2^k-1$, $k>1$. 
Finally, we establish that the independent domination number 
$\alpha_n\leq 3 \times 2^{n-k-2}$ for $7\times 2^{k-2}-1\leq n<2^{k+1}-1$, $k>1$. This is an improved upper bound for this range as compared to earlier work. 
\end{abstract}

 
\cfoot{\thepage\ of \pageref{LastPage}}


\section{Introduction and related work}
Hypercubes are classical structures that are well-studied in many important areas such as set theory, graph theory, combinatorics and Boolean algebra \cite{hays_harary}.
Given an $n$-dimensional hypercube graph $Q_n = (V_n,E_n)$, where $V_n$ be the set of vertices (nodes) and $E_n$ be the set of edges in $Q_n$, and a subset $S_n\subset V_n$, we say that $S_n$ dominates its neighbourhood $N(S_n)$, i.e., the set of all nodes that are adjacent to at least one node in $S_n$ \cite{harary,hays_harary}. If $S_n$ dominates $V_n\setminus S_n$, then $S_n$ is called a dominating set of $Q_n$. $S_n$ is said to be independent if no two nodes in $S_n$ are adjacent. An independent dominating set with minimum cardinality is called the minimum independent dominating set. The independent domination number $\alpha_n$ is the cardinality of the minimum independent dominating set of $Q_n$.\par
Many results are known on independent domination for simple regular and other graphs \cite{harvaland1}-\cite{survey}. However, there are many unsettled questions regarding minimum dominating sets in hypercubes as stated by Harary et al. \cite{harary}.
They derived the minimum independent dominating sets of $Q_n$ for dimension $n=1,2,3,4,5,6$:$\alpha_1=1, \alpha_2=2, \alpha_3=2, \alpha_4=4, \alpha_5=8, \alpha_6=12$. It was also reported that $\alpha_n=2^{n-k}$ for $n=(2^k-1)$ and $n=2^k$, $k>0$. 
Yet, the upper bound on $\alpha_n$ is still not tight for many other values of $n$. Mane et al. \cite{mane} showed that $\alpha_n\leq 2^{n-k}$ for $2^k<n<2^{k+1}-1$, for any $n$. In this work, we improve this bound for certain values of $n$. We split the range $2^k-1\leq n<2^{k+1}-1$ into two non-overlapping cases: \\\\
Case 1: $2^k-1\leq n< 7\times 2^{k-2}-1$, $k>1$;\\
Case 2: $7\times 2^{k-2}-1\leq n<2^{k+1}-1$, $k>1$.\\ 
\par
In this paper, we show that 
$\alpha_n\leq 3\times 2^{n-k-2}$ for $7\times 2^{k-2}-1\leq n<2^{k+1}-1,k>1$. This tightens the upper bound for Case 2 in comparison to the result $\alpha_n\leq 2^{n-k}$ proved earlier for both cases by Mane et al. \cite{mane}. Given an independent dominating set $S_p$, our procedure constructs an independent dominating set  $S_n$, where $|S_n|=2^p \times |S_p|$, $n=2p+1$, $p$ being a positive integer $\ge 1$.
Our result follows from a procedure that iteratively constructs a minimal independent dominating set in a hypercube from that of a smaller-dimension hypercube. 
\section{Concepts and techniques}\label{double-concat}\label{concept}
The $n$-dimensional hypercube $Q_n = (V_n,E_n)$ is an undirected regular graph whose vertex set where $V_n$ and $E_n$ denote the set of nodes and set of edges in $Q_n$, respectively. 
The set $V_n$ can be envisaged as the set of all binary $n$-tuples of zeros and ones, i.e. $\{0,1\}^n$. Thus, $|V_n|=2^n$. Two nodes of $Q_n$ are adjacent if their binary $n$-tuples differ in exactly one place, i.e., if they are unit Hamming distance apart. Let the Hamming distance between two binary $n$-tuples corresponding to vertices $v_1,v_2\in V_n$ be represented as $d_H(v_1,v_2)$. 
\begin{lem}\label{lower_bound}
$\alpha_n \ge \lfloor(2^n/n+1)\rfloor$
\end{lem}
\textbf{\textit{Proof :}}
For any $Q_n=(V_n,E_n)$, $|V_n|=2^n$ and every vertex in $V_n$ has $n$ adjacent vertices. That is, a vertex of $V_n$ can dominate at most $n$ vertices uniquely. Thus, the minimum number of vertices required to dominate all $2^n$ vertices of $Q_n$ will be $2^n/(n+1)$. As the number of vertices should be an integer, $\alpha_n > \lfloor(2^n/n+1)\rfloor$.
\hfill \fbox{}
\section{Iterative construction of an independent dominating set in a hypercube}
The following procedure constructs an independent dominating set $S_{n+1}$ of $Q_{n+1}$, given an independent dominating set $S_n$ of $Q_n$, where $|S_{n+1}|=2\times|S_n|$. We use the following notation in Procedure 1.
\par
Let $S'_n$ be an independent dominating set of $Q_n$ such that $|S'_n|=|S_n|$. Let a vertex $v_{k,n}\in S_n$, $v'_{k,n}\in S'_n$ be represented by $n$-bit binary vectors $b^k_{n-1}b^k_{n-2}\cdots b^k_1b^k_0$, $b'^k_{n-1}b'^k_{n-2}\cdots b'^k_1b'^k_0$, respectively, where $b^k_l\in \{0,1\}$, $b'^k_l\in \{0,1\}$, $(1\leq k\leq |S_n|)$, $ (0\leq l<n)$.\\\\
{\bf{Procedure 1: }}\\
Input: $S_n$\\
Output: $S_{n+1}$
\begin{enumerate}[Step 1.]
\item Assign $S'_n \leftarrow \phi, S_{n+1} \leftarrow \phi$.
\item For every vertex $v_{k,n}\in S_n$ get a vertex $v'_{k,n}$, such that $v_{k,n},v'_{k,n}$ differ in LSB, i.e., $v_{k,n}=b^k_{n-1}b^k_{n-2}\cdots b^k_1b^k_0$, and $v'_{k,n}=b'^k_{n-1}b'^k_{n-2}\cdots b'^k_1b'^k_0=b^k_{n-1}b^k_{n-2}\cdots b^k_1\overline{b^k_0}$.
\item $\forall v'_{k,n}$, $S'_n=S'_n \cup \{v'_{k,n}\}$.
\item For every $v_{k,n}\in S_n$, and for every $v'_{k,n}\in S'_n$ get $v_{k,n+1}, v'_{k,n+1}$ respectively, such that $v_{k,n+1}=b^k_n.v_{k,n}=b^kb^k_{n-1}b^k_{n-2}\cdots b^k_1b^k_0$, $b^k_n\in \{0,1\}$, and $v'_{k,n+1}=b'^k_n.v'_{k,n}=b'^kb'^k_{n-1}b'^k_{n-2}\cdots b'^k_1b'^k_0=\overline{b^k_n}b^k_{n-1}b^k_{n-2}\cdots b^k_1\overline{b^k_0}$, $b'^k_n\in \{0,1\}$.\\
\item $S_{n+1}=S_{n+1} \cup \{v_{k,n+1},v'_{k,n+1}\}, \forall v_{k,n}\in S_n, \forall v'_{k,n}\in S'_n$.
\item Output $S_n$.
\item End.
\end{enumerate}
\begin{lem}\label{twice_cardinality}
$|S_{n+1}|=2 \times |S_n|$
\end{lem}
\textbf{\textit{Proof :}} \textnormal{It follows from Procedure 1}.
\hfill \fbox{}
\begin{lem}
$S_{n+1}$ as constructed by Procedure 1 is an independent dominating set of $Q_n$.
\end{lem}	
\textbf{\textit{Proof :}} Follows from \cite{discretejournal}.\hfill \fbox{}
\section{Constructing an independent dominating set for $Q_n$, where $n=(2p+1),p\ge 1$}
\begin{figure}[]
\begin{mdframed}
\begin{mdframed}
\begin{tikzpicture}
\node[rectangle,draw]  at (60,29) (5){$N_1$};
    \node[rectangle,draw] at (65,29)(6) {$A^{N_1}=A^{N_1}_{p-1}A^{N_1}_{p-2}\cdots A^{N_1}_{0}$};
    \node[rectangle,draw]  at (60,28) (7){$6$};
    \node[rectangle,draw] at (65,28)(8) {$A^6=A^6_2A^6_1A^6_0=110$};
    \begin{scope}
      \path[->] (5) edge node {} (6);
      \path[->] (7) edge node {} (8);
	\end{scope}
  \end{tikzpicture}
 \end{mdframed}
 \begin{mdframed}
  \begin{tikzpicture}
   \node[rectangle,draw]  at (60,27) (13){$s^p_{k_1}$};
 \node[rectangle,draw] at (65,27)(14) {$b^{k_1}_{p-1}b^{k_1}_{p-2} \cdots b^{k_1}_0$};
    \node[rectangle,draw]  at (60,26) (15){$s^3_{1}$};
    \node[rectangle,draw] at (65,26)(16) {$b^1_2b^1_1b^1_0=010$};
    \begin{scope}
      \path[->] (13) edge node {} (14);
      \path[->] (15) edge node {} (16);
	\end{scope}
    \end{tikzpicture}
  \end{mdframed}
 \begin{mdframed}
\begin{tikzpicture}
\node[rectangle,draw] at (60,31)(9) {Parity of $A^{N_1}$};
 \node[rectangle,draw]  at (65,31) (10){$c^{N_1 \times |S_p|+k_1}_{n-1}=c^{l_1}_{n-1}=c^{l_1}_{2p}$};
 \node[rectangle,draw] at (60,30)(11) {Parity of $A^6$};
        \node[rectangle,draw]  at (65,30) (12){$c^{13}_6=0$};
        \begin{scope}
      \path[->] (9) edge node {} (10);
      \path[->] (11) edge node {} (12);
	\end{scope}
  \end{tikzpicture}
  \end{mdframed}
    \begin{mdframed}
\begin{tikzpicture}
\node[rectangle,draw]  at (60,27) (e){$A^{N_1}$};
 \node[rectangle,draw] at (65,27)(f) {$c^{l_1}_{n-1}c^{l_1}_{n-2}\cdots c^{l_1}_0=c^{l_1}_{2p-1}c^{l_1}_{2p-2}\cdots c^{l_1}_0$};
  \node[rectangle,draw] at (60,26)(g) {$A^6$};
         \node[rectangle,draw]  at (65,26) (h){$c^{13}_5c^{13}_4c^{13}_3=010$};
       \begin{scope}
      \path[->] (e) edge node {} (f);
      \path[->] (g) edge node {} (h);
	\end{scope}
    \end{tikzpicture}
  \end{mdframed}
  \begin{mdframed}
\begin{tikzpicture}
\node[rectangle,draw] at (60,21)(a) {$\overline{b^1_2}\overline{b^1_1}{b^1_0}$};
  \node[rectangle,draw]  at (65,21) (b){$c^{l_1}_{p-1}c^{l_1}_{p-2}\cdots c^{l_1}_0$};
     \node[rectangle,draw] at (60,20)(c) {$100$};
    \node[rectangle,draw]  at (65,20) (d){$c^{13}_2c^{13}_1c^{13}_0$};
       \begin{scope}
      \path[->] (a) edge node {} (b);
      \path[->] (c) edge node {} (d);
	\end{scope}
    \end{tikzpicture}
  \end{mdframed}
  \begin{mdframed}
\begin{tikzpicture}
  \node[rectangle,draw]  at (60,15) (21){$s^n_{N\times |S_p|+k_1}$};
    \node[rectangle,draw] at (65,15)(22) {$c^{l_1}_{n-1}c^{l_1}_{2p-1}c^{l_1}_{2p-2}\cdots c^{l_1}_{p-1}c^{l_1}_{p-2}\cdots c^{l_1}_0$};
    \node[rectangle,draw]  at (60,14) (23){$s^7_{0\times 2+1}=s^7_{13}$};
    \node[rectangle,draw] at (65,14)(24) {$c^{13}_{6}c^{13}_{5}c^{13}_{4}\cdots c^{13}_{3}c^{13}_{2}c^{13}_0=0110100$};  
    \begin{scope}
      \path[->] (21) edge node {} (22);
      \path[->] (23) edge node {} (24);
	\end{scope} 
    \end{tikzpicture}
      \end{mdframed}
      \end{mdframed}
    \caption{Construction of an $n=(2p+1)$-dimensional vector of the independent dominating set $S_n$ from a $p$-dimensional vector of the dominating set $S_p$, where $n=(2p+1)$. It illustatrates the construction of a vector for $p=3$, $S_3=\{010,101\}$, i.e., the construction of a vector of $S_7$ from a vector of $S_3$.}
\label{from s_3 to s_7}
\end{figure}
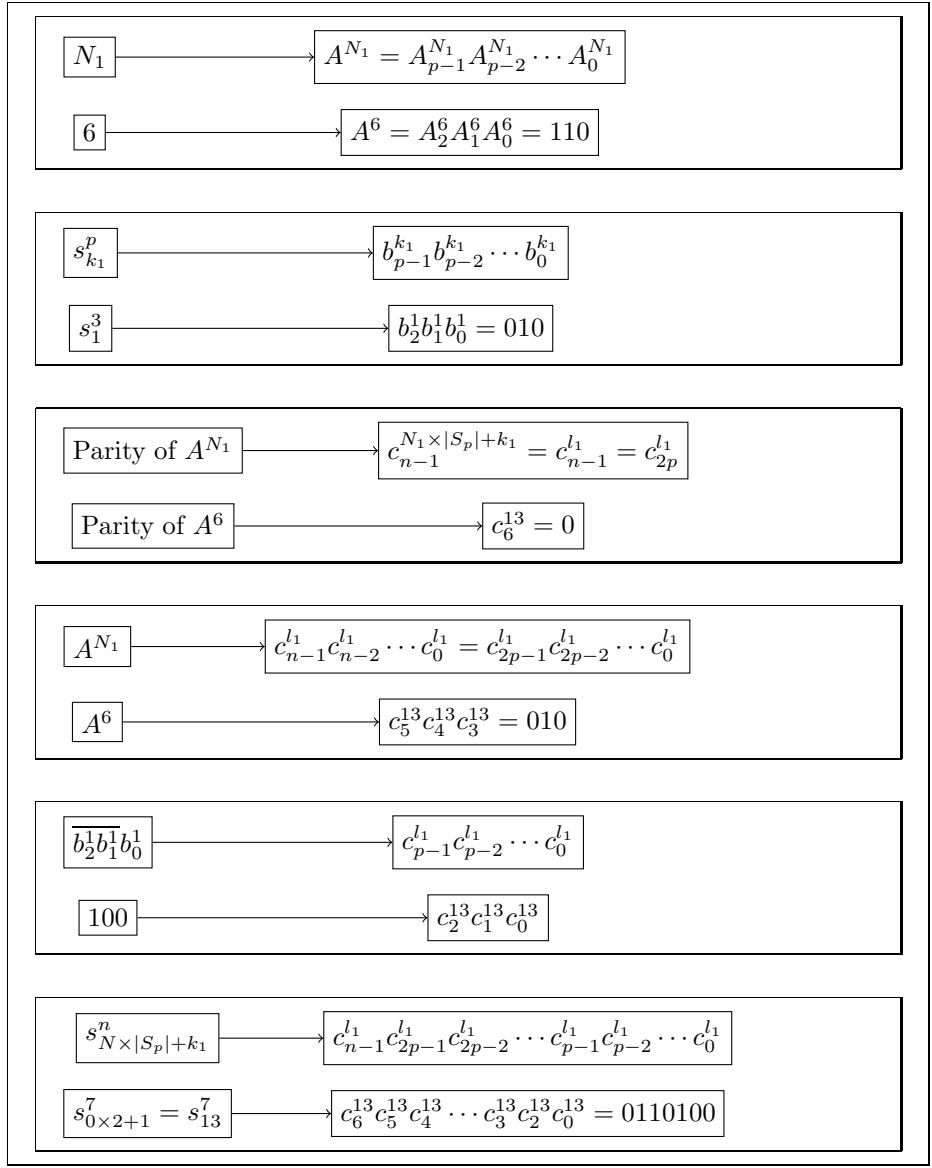

Let $S_p,S_n$ be independent dominating sets for $p$-, and $n$-dimensional hypercubes $Q_p=(V_p,E_p)$ and $Q_n=(V_n,E_n)$, respectively, where $n=2p+1$. Clearly, $S_p\subset V_p$, $S_n\subset V_n$, $|V_p|=2^p$, $|V_n|=2^n$. 
Let $s^{p}_{k}$, $s^{n}_{l}$ be vertices of $Q_p$ and $Q_n$, respectively, such that $s^p_k\in S_p$, $(1\leq k\leq |S_p|)$, $s^n_l\in S_n$, $(1\leq l\leq |S_n|)$. 
The following procedure constructs $S_n$ from $S_p$.\\\\
{\bf Procedure 2:}\label{algo}\\
Input: $S_p$ where every vector $s^p_k\in S_p$ is represented as $p$-bit binary $s^p_k=b^k_{(p-1)}b^k_{(p-2)}\cdots b^k_i\cdots b^k_1 b^k_{0}$, where $b^k_i\in \{0,1\}, (0\leq i\leq p-1)$.\\
Output: $S_n$ where every vector $s^n_l\in S_n$ is represented as $n$-bit binary $s^n_l=c^l_{(n-1)}c^l_{(n-2)}\cdots c^l_i\cdots c^l_1 c^l_{0}$, where $c^l_i\in \{0,1\}, (0\leq i\leq n-1)$.
\begin{enumerate}[Step 1.]
\item Assign $N\leftarrow 0$, $S_n\leftarrow \phi$
\item Get $p$-bit binary tuple of $N$ as $A^N=a^N_{p-1}a^N_{p-2}\cdots a^N_i\cdots a^N_0$, where $a^N_i\in\{0,1\}$, $(0\leq i\leq p-1)$.
\item Set $k=1$.
\item Set $l=N \times |S_p|+k$.
\item $j \leftarrow 0$.
\item $c^l_{j+p}=a^N_j$.
\item If $a^N_j=1$, then $c^l_{j}=\overline{b^k_j}$ else $c^l_j=b^k_j$.
\item $j=j+1$; If $j<p$ then goto Step 6.
\item If $A^N$ is of even parity then $c^l_{n-1}=0$ else $c^l_{n-1}=1$.
\item $S_n=S_n\cup \{s^n_l\}$.
\item $k=k+1$; If $k\leq |S_p|$, then goto Step 4.
\item $N=N+1$; If $N < 2^p$, then goto Step 2.
\item Output $S_n$.
\item End.
\end{enumerate}
\begin{exa}
Table 1 shows how Procedure 2 is executed, given a minimum independent dominating set $S_3=\{000,111\}$ for the 3-dimensional hypercube $Q_3$ to obtain an independent dominating set $S_7$ for the 7-dimensional hypercube $Q_7$. Also, set $S_7$ produced by Procedure 2 is a minimum independent dominating set in $Q_7$. Figure \ref{from s_3 to s_7}
shows how a vector of $S_3$ is modified to produce a vector of $S_7$. It also includes, in general, the concatenation process that leads to an $n=(2p+1)$-dimensional vector of the dominating set $S_n$ starting from a $p$-dimensional vector of the dominating set $S_p$. 
Given, $p=3$, $n=7$, $1\leq k\leq |S_p|$, i.e., $1\leq k\leq 2$, $k=\{k_1,k_2\}=\{1,2\}$ and $S_p=\{s^p_{k_1},s^p_{k_2}\}=\{s^3_1,s^3_2\}=\{010,101\}$.
\end{exa}
\begin{lem}\label{odd_lemma1}
$S_n$ as produced by Procedure 2 has cardinality $|S_n|=2^p \times |S_p|$.
\end{lem}
\textbf{\textit{Proof : }}
\textnormal{
Immediate from Procedure 2.} \hfill \fbox{}
\begin{theo}\label{odd_lemma2}
$S_n$ as obtained by Procedure 2, is an independent dominating set of $Q_n$.
\end{theo}
\textit{\textbf{Proof : }}
\textnormal{
First, we prove that $S_n$ is an independent set of $Q_n$. Next, we will show that $S_n$ is also a dominating set of $Q_n$.\\
Let $s^n_{l_1}, s^n_{l_2}\in S_n$ be two vertices of $Q_n$ obtained by Procedure 2.
Clearly, 
\begin{align*}
& s^n_{l_1}=c^{l_1}_{n-1}c^{l_1}_{n-2}c^{l_1}_{n-3}\cdots c^{l_1}_{p}c^{l_1}_{p-1}c^{l_1}_{p-2}\cdots c^{l_1}_{0}\nonumber\\
& =c^{l_1}_{n-1}a^{N_1}_{p-1}a^{N_1}_{p-2}\cdots a^{N_1}_{0}b^{k_1^*}_{p-1}b^{k_1^*}_{p-2}\cdots b^{k_1^*}_{0} \text{, where either } b^{k_1^*}_{p-1}=b^{k_1}_{p-1} \text{ or } \overline{b^{k_1}_{p-1}}\nonumber\\
& \text{and}\nonumber\\
& s^n_{l_2}=c^{l_2}_{n-1}c^{l_2}_{n-2}c^{l_2}_{n-3}\cdots c^{l_2}_{p}c^{l_2}_{p-1}c^{l_2}_{p-2}\cdots c^{l_2}_{0}\nonumber\\
& =c^{l_2}_{n-1}a^{N_2}_{p-1}a^{N_2}_{p-2}\cdots a^{N_2}_{0}b^{k_2^*}_{p-1}b^{k_2^*}_{p-2}\cdots b^{k_2^*}_{0} \text{, where either } b^{k_2^*}_{p-1}=b^{k_2}_{p-1} \text{ or } \overline{b^{k_2}_{p-1}}, \nonumber\\
& (1\leq l_1,l_2\leq |S_n|), |S_n|=2^p \times |S_p|, l_1\neq l_2, 0\leq N_1,N_2\leq (2^p-1), 1\leq k_1,k_2\leq |S_p|\nonumber
\end{align*}
For any two vetices $s^{n}_{l_1}, s^n_{l_2}\in S_n$, there are three possible cases;\\
Case 1: $N_1=N_2$; \\
Case 2: $N_1\neq N_2$, $c^{l_1}_{n-1}=c^{l_2}_{n-1}$, i.e., both $A^{N_1},A^{N_2}$ have the same parity, either both even or both odd;\\
Case 3: $c^{l_1}_{n-1} \neq c^{l_2}_{n-1}$, i.e., if $A^{N_1}$ has even (odd) parity, then $A^{N_2}$ has odd (even) parity.\\
\indent
For Case 1, the Hamming distance between $n$-bit binary vectors of $s^n_{l_1}, s^n_{l_2}$ is 
\begin{align*}
& d_H(s^n_{l_1},s^n_{l_2})\nonumber\\
& =d_H(c^{l_1}_{n-1},c^{l_2}_{n-1})+d_H(A^{N_1},A^{N_2})+d_H(b^{k_1^*}_{p-1}b^{k_1^*}_{p-2}\cdots b^{k_1^*}_{0},b^{k_2^*}_{p-1}b^{k_2^*}_{p-2}\cdots b^{k_2^*}_{0})\nonumber\\
& =(d_H(c^{l_1}_{n-1},c^{l_2}_{n-1})+d_H(A^{N_1},A^{N_2})+d_H(s^p_{k_1},s^p_{k_2}))\geq 2\nonumber\\
& \text{Note that } d_H(c^{l_1}_{n-1},c^{l_2}_{n-1})=d_H(A^{N_1},A^{N_2})=0 \text{, and } d_H(s^p_{k_1},s^p_{k_2})\geq 2 \text{ as } s^p_{k_1},s^p_{k_2}\in S_p\nonumber
\end{align*}
For Case 2,
\begin{align*}
& d_H(s^n_{l_1},s^n_{l_2})\nonumber\\
& =(d_H(c^{l_1}_{n-1},c^{l_2}_{n-1})+d_H(A^{N_1},A^{N_2})+d_H(b^{k_1^*}_{p-1}b^{k_1^*}_{p-2}\cdots b^{k_1^*}_{0},b^{k_2^*}_{p-1}b^{k_2^*}_{p-2}\cdots b^{k_2^*}_{0}))\geq 2\nonumber\\
& \text{when } d_H(c^{l_1}_{n-1},c^{l_2}_{n-1})=0, \text{ } d_H(A^{N_1},A^{N_2})\geq 2\nonumber
\end{align*}
For Case 3,
\begin{align*}
& d_H(s^n_{l_1},s^n_{l_2})\nonumber\\
& =(d_H(c^{l_1}_{n-1},c^{l_2}_{n-1})+d_H(A^{N_1},A^{N_2})+d_H(b^{k_1^*}_{p-1}b^{k_1^*}_{p-2}\cdots b^{k_1^*}_{0},b^{k_2^*}_{p-1}b^{k_2^*}_{p-2}\cdots b^{k_2^*}_{0}))\geq 2\nonumber\\
& \text{when } d_H(c^{l_1}_{n-1},c^{l_2}_{n-1})=1, \text{ } d_H(A^{N_1},A^{N_2})\geq 1\nonumber
\end{align*}
Thus, for any two vertices $s^n_{l_1}, s^n_{l_2}\in S_n$, we have $d_H(s^n_{l_1},s^n_{l_2})\geq 2$. That is, no two vertices in $S_n$ are adjacent. Hence, $S_n$ is an independent set of $Q_n$.}\par
To prove that $S_n$ is a dominating set, we argue as follows. We have proved that $S_n$ is an independent set, i.e., its no two vertices are adjacent, i.e., $\forall v_1,v_2\in S_n$,  $d_H(v_1,v_2)\ge 2$. Thus, for every vertex $v_n\in S_n$, its all $n$ adjacent vertices are in $S'_n$. Thus, we have the following four cases:\\
Case 1: $S'_n$ contains only $N(S_n)$;\\
Case 2: $S_n\cap S'_n=\phi$;\\
Case 3: $\forall v'_n\in S'_n$, $\exists $ a $v_n\in S_n$, such that $v'_n$ is adjacent to $v_n$.\par
Let $N(v_1), N(v_2)$ be the neighbourhood of $v_1,v_2\in S_n$, respectively. Now, $\forall v_1, v_2\in S_n$, if every $N(v_1)\cap N(v_2)=\phi$, then 
\begin{align*}
&|S_n|+|S'_n|\nonumber\\
&=(2^p\times |S_p|) + (n \times 2^p \times |S_p|)\nonumber\\
&=2^p \times |S_p| \times (n+1)\nonumber\\
&=|S_p| \times 2^p \times (2p+2)\nonumber\\
&=2^{p+1} \times (p+1) \times |S_p|\nonumber\\
&=2^{p+1}\times (|S_p|+p \times |S_p|)\nonumber\\
&=2^{p+1}\times (|S_p|+|S'_p|)\nonumber\\
&=2^{p+1}\times 2^p=2^{2p+1}=2^n\nonumber\\
&\text{So, }\nonumber\\
&|S_n|+|S'_n|=|V_n|\nonumber
\end{align*}
Cace 4: $S_n \cup S'_n=V_n$.\\
By combining Cases (1), (2) and (4), we have $S'_n=V_n \setminus S_n$, i.e., $S'_n$ contains all those vertices of $Q_n$ which are not in $S_n$. Again, from Case 3, $\forall v'_n\in S'_n, \exists $ a $v_n\in S_n$, such that $v'_n$ is adjacent to $v_n$. Thus, $S_n $ is a dominating set of $Q_n$. 
Hence, $S_n$ is an independent dominating set of $Q_n$. \hfill \fbox{}
\par
Procedure 2 leads to the following known result \cite{harary}. 
\begin{lem}
$\alpha_n=2^{n-k}$ for $n=2^k-1,k>0$.
\end{lem}
\textit{\textbf{Proof: }} \textnormal{
From Lemma \ref{lower_bound}, $\alpha_n\ge \lfloor(2^n/n+1)\rfloor$. From Lemma \ref{lower_bound}, it follows that $\alpha_n>\lfloor(2^n/n+1)\rfloor$ . When $n=2^k-1$, $2^n/(n+1)=2^{n-k}$ is an integer, and hence,  $\alpha_n=2^{n-k}$.\hfill \fbox{}
}
\begin{table}[]
\begin{tabular}{|l|l|l|l|l|l|l|l|}
\hline
$N$ & $A^N$ & $k$ & $l$ &  $c^l_{n-1}$ & $s^p_k=s^3_k$ & $c^l_{p-1}c^l_{p-2}\cdots c^l_{0}$ & $s^n_{|S_p|\times N+k}$\\
 & $=a^N_{p-1}a^N_{p-1}\cdots a^N_{0}$ &  &  & $=c^l_{6}$  & $=b^k_{p-1}b^k_{p-2}\cdots b^k_{0}$ & $=c^l_{2}c^l_{1}c^l_{0}$ & $=s^7_{2\times N+k}$\\
 & $=a^N_{2}a^N_{1}a^N_{0}$ &  &  & (parity & $=b^k_{2}b^k_{1}b^k_{0}$ & &$=c^l_{n-1}A^Nc^l_{p-1}$\\
& & & & of $A^N)$ & & & $\cdots c^l_1c^l_{0}$\\
\hline
0 & \textcolor{green}{000} & 1 & 1 &  \textcolor{red}{0} & 000 & \textcolor{blue}{000} & \textcolor{red}{0}\textcolor{green}{000}\textcolor{blue}{000}\\
\hline
0 & \textcolor{green}{000} & 2 & 2 & \textcolor{red}{0} & 111 & \textcolor{blue}{111} & \textcolor{red}{0}\textcolor{green}{000}\textcolor{blue}{111}\\
\hline
1 & \textcolor{green}{001} & 1 & 3 &  \textcolor{red}{1} & 000 & \textcolor{blue}{001} & \textcolor{red}{1}\textcolor{green}{001}\textcolor{blue}{001}\\
\hline
1 & \textcolor{green}{001} & 2 & 4  & \textcolor{red}{1} & 111 & \textcolor{blue}{110} & \textcolor{red}{1}\textcolor{green}{001}\textcolor{blue}{110}\\
\hline
2 & \textcolor{green}{010} & 1 & 5 & \textcolor{red}{1} & 000 & \textcolor{blue}{010} & \textcolor{red}{1}\textcolor{green}{010}\textcolor{blue}{010}\\
\hline
2 & \textcolor{green}{010} & 2 & 6 & \textcolor{red}{1} & 111 & \textcolor{blue}{101} & \textcolor{red}{1}\textcolor{green}{010}\textcolor{blue}{101}\\
\hline
3 & \textcolor{green}{011} & 1 & 7 & \textcolor{red}{0} & 000 & \textcolor{blue}{011} & \textcolor{red}{0}\textcolor{green}{011}\textcolor{blue}{011}\\
\hline
3 & \textcolor{green}{011} & 2 & 8 & \textcolor{red}{0} & 111 & \textcolor{blue}{100} & \textcolor{red}{0}\textcolor{green}{011}\textcolor{blue}{100}\\
\hline
4 & \textcolor{green}{100} & 1 & 9 & \textcolor{red}{1} & 000 & \textcolor{blue}{100} & \textcolor{red}{1}\textcolor{green}{100}\textcolor{blue}{100}\\
\hline
4 & \textcolor{green}{100} & 2 & 10 & \textcolor{red}{1} & 111 & \textcolor{blue}{011} & \textcolor{red}{1}\textcolor{green}{100}\textcolor{blue}{011}\\
\hline
5 & \textcolor{green}{101} & 1 & 11 & \textcolor{red}{0} & 000 & \textcolor{blue}{101} & \textcolor{red}{0}\textcolor{green}{101}\textcolor{blue}{101}\\
\hline
5 & \textcolor{green}{101} & 2 & 12 & \textcolor{red}{0} & 111 & \textcolor{blue}{010} & \textcolor{red}{0}\textcolor{green}{101}\textcolor{blue}{010}\\
\hline
6 & \textcolor{green}{110} & 1 & 13 & \textcolor{red}{0} & 000 & \textcolor{blue}{110} & \textcolor{red}{0}\textcolor{green}{110}\textcolor{blue}{110}\\
\hline
6 & \textcolor{green}{110} & 2 & 14 & \textcolor{red}{0} & 111 & \textcolor{blue}{001} & \textcolor{red}{0}\textcolor{green}{110}\textcolor{blue}{001}\\
\hline
7 & \textcolor{green}{111} & 1 & 15 & \textcolor{red}{1} & 000 & \textcolor{blue}{111} & \textcolor{red}{1}\textcolor{green}{111}\textcolor{blue}{111}\\
\hline
7 & \textcolor{green}{111} & 2 & 16 & \textcolor{red}{1} & 111 & \textcolor{blue}{000} & \textcolor{red}{1}\textcolor{green}{111}\textcolor{blue}{000}\\
\hline
\end{tabular}
\caption{Construction of $S_7$ from $S_3$}
\label{table_construct}
\end{table}

\section{Upper bound on the independent domination number for hypercubes}
\begin{theo}\label{alpha_value22}
For $n=7 \times 2^{k-2}-1$, where $k>1$,  $\alpha_n\leq 3 \times 2^{n-k-2}$
\end{theo}
\textbf{\textit{Proof : }}
From \cite{harary}, we know that $\alpha_6=12$. In this case, $k=2$. Following Lemma \ref{odd_lemma1} and Theorem \ref{odd_lemma2}, for $n=13$, we obtain $|S_{13}|=12 \times 2^6$, where $k=3$. By iteration, we can then derive results for $n=27, 55, 111, 223,  \cdots,$ corresponding to $k=4, 5, 6, 7, \cdots$, respectively. Following Procedure 2, we obtain $|S_n|=12 \times 2^6,12 \times 2^{19},12 \times 2^{46},\cdots$ for $n=13, 27, 55, \cdots$, respectively. The $k^{th}$ term in the series can be evaluated as $7 \times 2^{k-2}-1$. Hence, the proof. 
\hfill \fbox{}
\begin{theo}\label{alpha_value2}
For $7 \times 2^{k-2}-1\leq n< 2^{k+1}-1$, where $k>1$,  $\alpha_n\leq 3 \times 2^{n-k-2}$
\end{theo}
\textbf{\textit{Proof : }}
If $S_n$ is given, we can construct $S_{n+1}$ following Procedure 1. Thus, when $|S_n|=3 \times 2^{n-k-2}$, $|S_{n+1}|=2 \times |S_n|=3 \times 2^{(n+1)-k-2}$. Hence, the proof. 
\hfill \fbox{}
\par
Thus it improves the previous upper bound $\alpha_n\leq 2^{n-k}$ \cite{mane} for the range of $7 \times 2^{n-k}\leq n<2^{k+1}-1$. 
\begin{theo}\label{alpha_value3}
For $2^{k}-1\leq n< 7 \times 2^{k-2}-1$, $k>0$, $\alpha_n\leq 2^{n-k}$.
\end{theo}
\textbf{\textit{Proof : }}
If $S_n$ is given, we can construct $S_{n+1}$ following Procedure 1. Thus, when $|S_n|=2^{n-k}$, $|S_{n+1}|=2 \times |S_n|=2^{(n+1)-k}$. Hence, the proof. 
\hfill \fbox{}
\section{Independent domination number for hypercubes}
Here, we summarize our results and compare them with previously known bounds.
\subsection{Previous results}
For $n=2^k-1$, $k\geq 1$, $\alpha_n=2^{n-k}$ \cite{harary,thompson,berlekamp,lee} \\
For $n=2^k$, $k\geq 1$, $\alpha_n=2^{n-k}$ \cite{harary,mane,wee} \\
For $2^k<n<2^{k+1}-1$, $k\geq 1$, $\alpha_n\leq 2^{n-k}$ \cite{mane} 
\subsection{Proposed work}
For $n=2^{k}-1$, $k>0$, $\alpha_n=2^{n-k}$\\
For $7 \times 2^{k-2}-1\leq n < 2^{k+1}-1$, $k>1$, $\alpha_n\leq 3 \times 2^{n-k-2}$\\
For $2^{k}-1\leq n < 7 \times 2^{k-2}-1$, $k>0$, $\alpha_n\leq 2^{n-k}$\\
\subsection{Numerical comparison for some values of $n$}
We report the values of $\alpha_n$ for different values of $n$ in Table 2. 
Results in Table 3 are obtained from Table 2, which show improved upper bound for some values on $n\le 62$. 
\begin{table}[H]
\begin{tabular}{|c|c|c|c|c|}
\hline
& \multicolumn{2}{c}{$k>0$} &  & $k>1$ \\
\hline
$n$& $2^k-1$ & $2^k$ & $2^k<n< 7 \times 2^{k-2}-1$ & $7 \times 2^{k-2}-1\leq n<2^{k+1}-1$  \\
\hline
$\alpha_n$ \cite{harary,mane, thompson,berlekamp,lee,wee} & $2^{n-k}$ & $2^{n-k}$ & & \\
\hline
$\alpha_n$ \cite{mane} & & $2^{n-k}$ & $\leq 2^{n-k}$ & $\leq 2^{n-k}$   \\
\hline
$\alpha_n$ (this work) & $2^{n-k}$ & $\leq 2^{n-k}$ & $\leq 2^{n-k}$  & $\leq 3 \times 2^{n-k-2}$  \\
\hline
\end{tabular}
\caption{$\alpha_n$ of $Q_n$ for $2^k-1\leq n\leq 2^{k+1}-1, k\geq 1$}
\end{table}
\begin{table}[H]
{
\begin{tabular}[h]{|c|c|c|c|c|}
\hline
$k$ & $n$ & $\alpha_n$ & $\alpha_n$ & $\alpha_n$\\
& & \cite{harary,mane, thompson} & \cite{mane} &  (this work)\\
& & \cite{berlekamp,lee} &  &  \\
& & \cite{wee,stanton} &  &  \\
\hline
\multirow{1}{*}{1}& 2 & 2 & $2$ & $\leq 2$ \\
\hline
\multirow{4}{*}{2} & 3 & 2 & & $2$\\
& 4 & 4 & $2^{2}$ & $\leq 2^{2}$\\
& 5 & 8 & $\leq 2^{3}$ & $\leq 2^{3}$ \\
& 6 & 12 & $\leq 2^{4}$ & $\leq 3 \times 2^{2}$\\
\hline
\multirow{8}{*}{3} & 7 & $2^{4}$  & & $2^{4}$ \\
& 8 & $2^{5}$ & $2^{5}$ & $\leq 2^{5}$\\
& 9 &  & $\leq 2^{6}$ & $\leq 2^{6}$\\
& 10 &  & $\leq 2^{7}$ & $\leq 2^{7}$\\
& 11 &  & $\leq 2^{8}$ & $\leq 2^{8}$\\
& 12 &   & $\leq 2^{9}$& $\leq 2^{9}$\\
& 13 &   & $\leq 2^{10}$& $\leq 3 \times 2^{8}$\\
& 14 &  & $\leq 2^{11}$ & $\leq 3 \times 2^{9}$\\
\hline
\end{tabular}
\hspace{2cm}
\begin{tabular}[h]{|c|c|c|c|c|}
\hline
$k$ & $n$ & $\alpha_n$ & $\alpha_n$ & $\alpha_n$\\
& & \cite{harary,mane, thompson} & \cite{mane} &  (this work)\\
& & \cite{berlekamp,lee} &  &  \\
& & \cite{wee,stanton} &  &  \\
\hline
\multirow{4}{*}{4}& 27 &  & $\leq 2^{23}$  & $\leq 3 \times 2^{21}$\\
& 28 &  & $\leq 2^{24}$  & $\leq 3 \times 2^{22}$\\
& 29 &   & $\leq 2^{25}$  & $\leq 3 \times 2^{23}$ \\
& 30 &   & $\leq 2^{26}$  & $\leq 3 \times 2^{24}$\\
\hline
\multirow{8}{*}{5}& 55 &  & $\leq 2^{50}$ & $\leq 3 \times 2^{48}$\\
& 56 &   & $\leq 2^{51}$ & $\leq 3 \times 2^{49}$\\
& 57 &  & $\leq 2^{52}$ & $\leq 3 \times 2^{50}$\\
& 58 &  & $\leq 2^{53}$ & $\leq 3 \times 2^{51}$\\
& 59 &  & $\leq 2^{54}$ & $\leq 3 \times 2^{52}$\\
& 60 &  & $\leq 2^{55}$ & $\leq 3 \times 2^{57}$\\
& 61 &   & $\leq 2^{56}$ & $\leq 3 \times 2^{54}$\\
& 62 &   & $\leq 2^{57}$ & $\leq 3 \times 2^{55}$\\
\hline
\end{tabular}
}
\label{table1to31}
\caption{$\alpha_n$ for $n\leq 62$}
\end{table}

\section{Conclusion}
In this work, we have derived an improved upper bound on independent domination number in an $n$-dimensional hypercube for certain range of values of $n$.


\end{document}